\documentclass[RNAAS]{aastex62}

\newcommand{\tess}{{\it TESS}}
\newcommand{\cofiam}{{\tt cofiam}}
\newcommand{\polyam}{{\tt polyam}}
\newcommand{\local}{{\tt local}}
\newcommand{\gp}{{\tt gp}}
\newcommand{\methmarg}{{\tt methmarg}}
\newcommand{\multinest}{{\sc MultiNest}}

\begin{document}

\title{Detection of the Occultation of 55 Cancri e with \textit{TESS}}

%% Note that the corresponding author command and emails has to come
%% before everything else. Also place all the emails in the \email
%% command instead of using multiple \email calls.
\correspondingauthor{David Kipping}
\email{dkipping@astro.columbia.edu}

%% The \author command can take an optional ORCID.
\author[0000-0002-4365-7366]{David Kipping}
\affiliation{Department of Astronomy, Columbia University,
550 W 120th St., New York NY 10027, USA}

\author{Tiffany Jansen}
\affiliation{Department of Astronomy, Columbia University,
550 W 120th St., New York NY 10027, USA}

%% Note that RNAAS manuscripts DO NOT have abstracts.

\begin{abstract}
55 Cancri e is an ultra-short period transiting Super-Earth observed by \tess\ in
Sector 21. Using this photometry, we measure the occultation depth in the
\tess\ bandpass, leveraging the precise transit light curve and comparing
multiple detrending methods. We measure the occultation depth to be
$(15.0\pm4.8)$\,ppm - a staggeringly small change in brightness, yet one
detected by \tess\ in just a single sector of data. This implies a brightness
temperature of $2800_{-160}^{+130}$\,K, which is around 1.5\,$\sigma$ greater
than expected given the mean depth measured with \textit{Spitzer}. This is
not a formally significant difference, and may be accounted for by the known
variability, or by an albedo of ${\sim}0.5$. In any case, future \tess\
observations of this system will provide an exciting opportunity to further
study this diminutive world's atmosphere.
\end{abstract}

%% Start the main body of the article. If no sections in the 
%% research note leave the \section call blank to make the title.
\section{Observations} 

55 Cancri e is a ${\sim}2$\,$R_{\oplus}$ transiting exoplanet \citep{winn:2011}
orbiting a naked-eye star with an ultra short-period of $0.74$\,days
\citep{dawson:2010}. The star was observed by \tess\ during Sector 21 of its
second year of observations. We downloaded the publicly available processed PDC
lightcurves \citep{jenkins:2016} and removed outliers with a moving median
filter and the photometric error flags. We detrended the photometry in
individual orbital period segments centered on the expected time of transit by
deweighting in-transit points and using four different standard algorithms:
\cofiam, \polyam, \local\ (a local linear regression) and \gp\ (a Gaussian
Process with a Mat\'ern 3/2 kernel). These resulting time series were then
combined by evaluating the median flux at each time stamp, and propagating the
error between methods into the new fluxes, dubbed \methmarg\ (see
\citealt{teachey:2018} for a detailed description of this and the individual
algorithms used). We then repeated this process, centering instead on the times
of occultation.

The 34 useable epochs present in the \methmarg\ for the transit light curve
were regressed using a \citet{mandel:2002} model with $q_1$-$q_2$ limb
darkening \citep{q1q2} coupled to \multinest\ \citep{feroz:2009}. This provided
a precise estimate of the ratio-of-radii ($p=0.01807_{-0.00016}^{+0.00014}$),
first-to-fourth contact duration ($T_{14} = 1.5499_{-0.0093}^{+0.0110}$\,hours)
and second-to-third contact duration ($T_{23} = 
1.486_{-0.010}^{+0.012}$\,hours). Since the occultation event is unaffected by
limb darkening, and the orbit is nearly circular \citep{dawson:2010}, then the
occultation will manifest as a trapezoidal-like decrease in brightness
half-a-period after the transit events. The depth of the occultation event
is unknown, $\delta_{\mathrm{occ}}$, as is the out-of-occultation baseline
flux level (although this is very close to unity due to the detrending),
but the two duration parameters are tightly constrained from the transit fits.
We can thus simply regress this trapezoidal model to the \tess\ occultation
photometry in order to constrain $\delta_{\mathrm{occ}}$.

We use a simple Metropolis MCMC algorithm to explore the two-dimensional
parameter volume, with uniform priors on the baseline level and
$\delta_{\mathrm{occ}}$. The trapezoid durations are drawn from Gaussians
approximating the transit posterior results. For the \methmarg\ occultation,
the marginalized posterior indicates $\delta_{\mathrm{occ}} =
(15.0\pm4.8)$\,ppm - thus indicating that the occultation event was indeed
positively detected. We repeated this on the four different versions of
the occultation light curve using the different detrending approaches.
\local\ and \polyam\ are in close agreement with each other (and not
surprisignly the method marginalized result), as shown in Figure~\ref{fig}. The
\cofiam\ method finds a depth nearly twice as large, whereas \gp\ finds one
${\sim}50$\% smaller. Thus, the presence of an occultation is broadly
recovered across the different methods, but the differences highlight how this
is at the limit of \tess' sensitivity, where detrending choices become
influential. Although the occultation was not detected by \textit{MOST} in a
similar bandpass, the 2\,$\sigma$ upper limit obtained by \citet{dragomir:2014}
is consistent with our measurement ($<35$\,ppm).

\section{Interpretation} 

\begin{figure*}
\begin{center}
\includegraphics[width=18.0cm,angle=0,clip=true]{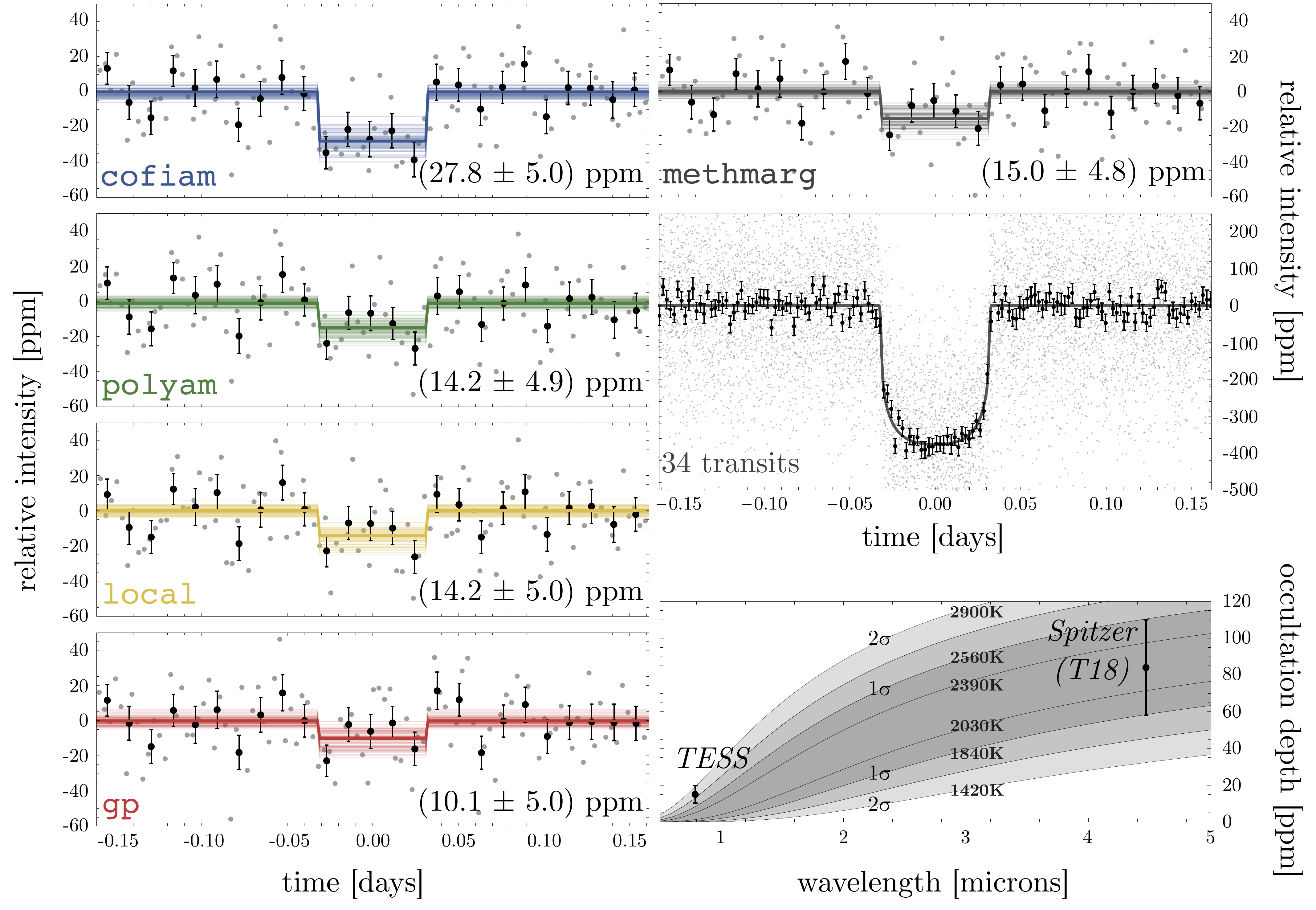}
\caption{\emph{
Left: Four different detrendings of the \tess\ photometry of 55 Cancri e,
surrounding the expected time of occultation. Right-top: Combination of
the four occultation light curves into a single ``method marginalized''
light curve, indicating positive evidence for an occultation event.
Right-middle: \tess\ transit photometry of 55 Cancri e, which we use
to learn the system parameters to high precision. Right-lower: Emission
spectrum of 55 Cancri e, plotting blackbody curves consistent with the
\textit{Spitzer} measurement of \citet{tamburo:2018}.
}}
\label{fig}
\end{center}
\end{figure*}

To briefly interpret the result, we integrated a blackbody curve for the star
multiplied by the \tess\ response function to solve for the brightness
temperature implied by our result. We find $2800_{-160}^{+130}$\,K, which
is larger than the equilibrium temperature expected of a pure blackbody at
55 Cancri e's orbital separation ($1970\pm15)$\,K). The occultation depth has
been previously shown to exhibit significant variability in the 4.5\,$\mu$m
band of \textit{Spitzer} \citep{tamburo:2018}, with brightness temperatures
of up to $(2700\pm250)$\,K. One possibility, then, is that \tess\ caught 55
Cancri e in an unusually bright episode in Sector 21. Another possibility is
that some reflected light is contributing to the occultation. For a geometric
albedo of unity, our transit posteriors indicate an expected reflectance
occultation depth of $(26\pm1)$\,ppm. The \citet{tamburo:2018} \textit{Spizter}
depths have a mean of $(84\pm26)$\,ppm, corresponding to
$2200_{-370}^{+350}$\,K after integrating over the 4.5 micron bandpass.
In Figure~\ref{fig}, we plot the TESS occultation depth where one can see that
the observed value is 1-2\,$\sigma$ higher than would be expected from a blackbody
curve derived from the brightness temperature of the \textit{Spitzer}
occultations. Although not formally ``significant'', it could
be explained by an albedo of ${\sim}0.5$, but we hesitate to place too much
confidence in that claim given the poorly understood nature of this planet's
variability. Nevertheless, future observations with \tess\ will provide
excellent opportunities to continue to monitor this planet's variability in the
post-\textit{Spitzer} era.

%% An example figure call using \includegraphics

%% RNAAS only allow one figure OR table, so this table is for the arXiv-only
%% version of the manuscript
\begin{deluxetable}{cc}
\tablecaption{One sigma credible intervals for the seven transit parameters
regressed to the \tess\ transit light curve of 55 Cnc e. \label{tab}}
\tablehead{
\colhead{Parameter} & \colhead{Value}
}
\startdata
$p$ & $0.01807_{-0.00016}^{+0.00014}$ \\
$\rho_{\star}$\,[g\,cm$^{-3}$] & $1.540_{-0.060}^{+0.061}$ \\
$b$ & $0.347_{-0.045}^{+0.039}$ \\
$P$\,[days] & $0.7365270_{-0.0000085}^{+0.0000086}$ \\
$\tau$\,[BJD-2570000] & $1882.47729_{-0.00011}^{+0.00010}$ \\
$q_1$ & $0.51_{-0.13}^{+0.16}$ \\
$q_2$ & $0.115_{-0.077}^{+0.118}$ \\
\enddata
\tablecomments{Fits assume a circular orbit.}
\end{deluxetable}

\section*{Acknowledgments}
\acknowledgments

This paper includes data collected by the TESS mission, which are publicly
available from the Mikulski Archive for Space Telescopes (MAST). Funding for
the TESS mission is provided by NASA’s Science Mission directorate. Special
thanks to Tom Widdowson, Mark Sloan, Laura Sanborn, Douglas Daughaday, Andrew
Jones, Jason Allen, Marc Lijoi, Elena West, Tristan Zajonc, Chuck Wolfred,
Lasse Skov \& Martin Kroebel.

\end{document}